\documentclass[12pt]{article}
\usepackage{amsmath,epsf,amssymb,latexsym,cite}
\usepackage[ps,dvips,matrix,arrow,frame,import,curve,color]{xy}

\newif\iffigs\figstrue

\DeclareFontFamily{U}{rsf}{}
\DeclareFontShape{U}{rsf}{m}{n}{
  <5> <6> rsfs5 <7> <8> <9> rsfs7 <10-> rsfs10}{}
\DeclareMathAlphabet\Scr{U}{rsf}{m}{n}

\def\pplogo{\vbox{\kern-\headheight\kern -29pt
\halign{##&##\hfil\cr&{%\sc
\ppnumber}\cr\rule{0pt}{2.5ex}&\ppdate\cr}
}}
\makeatletter
\def\ps@firstpage{\ps@empty \def\@oddhead{\hss\pplogo}%
  \let\@evenhead\@oddhead % in case an article starts on a left-hand page
}
\def\maketitle{\par
 \begingroup
 \def\thefootnote{\fnsymbol{footnote}}
 \def\@makefnmark{\hbox{$^{\@thefnmark}$\hss}}
 \if@twocolumn
 \twocolumn[\@maketitle]
 \else \newpage
 \global\@topnum\z@ \@maketitle \fi\thispagestyle{firstpage}\@thanks
 \endgroup
 \setcounter{footnote}{0}
 \let\maketitle\relax
 \let\@maketitle\relax
 \gdef\@thanks{}\gdef\@author{}\gdef\@title{}\let\thanks\relax}
\makeatother

\def\C{{\mathbb C}}
\def\P{{\mathbb P}}

\def\R{{\mathbb R}}
\def\Z{{\mathbb Z}}

\def\Vol{\operatorname{Vol}}

\def\Gr{\operatorname{Gr}}
\def\SO{\operatorname{SO}}
\def\Sl{\operatorname{SL}}
\def\Gl{\operatorname{GL}}
\def\GO{\operatorname{O{}}}
\def\SU{\operatorname{SU}}
\def\GU{\operatorname{U{}}}
\def\Sp{\operatorname{Sp}}

\def\rank{\operatorname{rank}}

\def\Spin{\operatorname{Spin}}

\def\Sym{\operatorname{Sym}}

\def\ch{\operatorname{\mathit{ch}}}
\def\td{\operatorname{\mathit{td}}}
\def\KIII{\mathrm{K3}}

\def\CY{Calabi--Yau}

\def\cM{{\Scr M}}

\def\cH{{\Scr H}}

\def\cF{{\Scr F}}

\def\ff#1#2{{\textstyle\frac{#1}{#2}}}

\def\nMM{n_{\textrm{M2}}}
\def\spnh{\Spin(32)/\Z_2}
\def\HS#1{{\mathbb{F}}_{#1}}

\begin{document}
\setcounter{page}0
\def\ppnumber{\vbox{\baselineskip14pt
\hbox{SU-ITP-05/13}
\hbox{DUKE-CGTP-05-02}
\hbox{hep-th/0504036}}}
\def\ppdate{April 2004} \date{}

\title{\LARGE An Analysis of Fluxes by Duality\\[10mm]}
\author{
Paul S.~Aspinwall\\[2mm]
\normalsize Department of Physics and SLAC \\
\normalsize Stanford University \\
\normalsize Stanford, CA 94305/94309\\
\normalsize and\\
\normalsize Center for Geometry and Theoretical Physics \\
\normalsize Box 90318 \\
\normalsize Duke University \\
\normalsize Durham, NC 27708-0318\\[8mm]
}

{\hfuzz=10cm\maketitle}

\def\Large{\large}
\def\LARGE{\large\bf}

\vskip 1cm

\begin{abstract}
M-theory on $\KIII\times\KIII$ with non-supersymmetry-breaking
$G$-flux is dual to M-theory on a \CY\ threefold times a 2-torus
without flux. This allows for a thorough analysis of the effects of
flux without relying on supergravity approximations. We discuss
several dual pairs showing that the usual rules of $G$-flux
compactifications work well in detail. We discuss how a transition can
convert M2-branes into $G$-flux. We see how new effects can arise at
short distances allowing fluxes to obstruct more moduli than one
expects from the supergravity analysis.
\end{abstract}

\vfil\break

%%%%%%%%%%%%%%%%%%%%%%%%%%%%%%%%%%%%%%%%%%%%%%%%%%%%%%%%%%%%%%%%

\section{Introduction}    \label{s:intro}

Flux compactifications along the lines of \cite{BB:8,GVW:8d,GKP:flux}
have received a good deal of attention recently. One aspect about flux
compactification that might make one a little uneasy is that the
analysis tends to depend on supergravity arguments. Effects due to
finite sizes which require the analysis of flux compactifications may
need to be treated more carefully.

In this paper we will undertake a rather random perusal of some
aspects of fluxes in the context of M-theory on
$\KIII\times\KIII$. This seems to be the most accessible nontrivial
case of fluxes. As we will see, there are vast possibilities even in
this case. In addition, we will show how there are inevitably some
fluxes which cannot be analyzed at large radius. These fluxes can fix
the volumes of the K3 surfaces.

The nice thing about M-theory on $\KIII\times\KIII$ with M2-branes or
$G$-flux that does not break any supersymmetry, is that it is dual to
a flux-free and brane-free compactification of M-theory on a \CY\
threefold, $X$, times a 2-torus. This allows us to check some aspects
of $G$-flux in a more rigorous context.

Our main plan is to follow extremal ``conifold''-like transitions in
the \CY\ threefold $X$ and see what happens in the dual
$\KIII\times\KIII$ picture. Some of the basic ideas of this analysis
are not new (see particularly \cite{BS:K3K3}, for example) but we try
to give a more complete picture of the interplay between the geometry
of $X$ and the fluxes on $\KIII\times\KIII$. One of the most useful
tools will be the ``stable degeneration'' picture of F-theory
\cite{FMW:F,AM:po}. The yields, for example, an explicit demonstration
of how the moduli of the K3 surfaces can be obstructed when $G$-flux is
turned on. Mirror symmetry also plays an interesting r\^ole
since it corresponds to exchanging the two K3 surfaces.

The $G$-flux will change as one passes through an extremal
transition. Since $G$ is in integral cohomology, the only way this can
happen is if the K3 surfaces become singular. This is indeed the
case as we shall see.

A full analysis of the complete web of \CY\ threefolds is still out of
reach since it involves, as yet, poorly-understood non-perturbative
effects. Thus, even in this simplest case, the subject of flux
compactification is highly nontrivial.

In section \ref{s:CYK} we will give several examples of how extremal
transitions are mapped into $G$-flux transitions. A particularly
interesting case concerns changing M2-branes into smooth $G$-flux
where the initial and final K3 surfaces are smooth. 

Section \ref{s:nonp} is more speculative in nature and concerns the
parts of the moduli space where nonperturbative effects become
important. We argue that new types of moduli obstructions can appear
because of fluxes and that the possibilities of M-theory
compactifications on $\KIII\times\KIII$ must exceed the number of
types of \CY\ threefolds. 

\section{\CY\ to K3$\times$K3 Dualities}  \label{s:CYK}

Let us recall some well-known general facts about M-theory
compactifications to three-dimensions with $N=4$ supersymmetry. Table
\ref{tab:hol} shows the various possibilities for how many
supersymmetries arise from the holonomy of a given manifold. The case
of $N=4$ is noteworthy since it is the highest value for $N$ which
arises in two different ways.

\begin{table}\begin{center}
\begin{tabular}{|c|c|c|}
\hline
$N$&Holonomy&Manifold\\
\hline
1&$\Spin(7)$&$\Spin(7)$-manifold\\
2&$G_2$&$G_2$-manifold $\times S^1$\\
2&$\SU(4)$&\CY\ fourfold\\
3&$\Sp(2)$&Hyperk\"ahler\\
4&$\Sp(1)\times \Sp(1)$&$\KIII\times\KIII$\\
4&$\SU(3)$&\CY $\times T^2$\\
8&$\Sp(1)$&$\KIII\times T^4$\\
16&1&$T^8$\\
\hline
\end{tabular}  
\caption{Three-dimensional compactifications of M-theory.}\label{tab:hol}
\end{center}\end{table}

Suppose we are given an M-theory compactification on $S_1\times S_2$,
where $S_1$ and $S_2$ are K3 surfaces. M-theory on $S_1$ is known
\cite{W:dyn} to be dual to the heterotic string on $T^3$. The
heterotic string on $T^2\times S_2$ is frequently dual to the type IIA
string on a \CY\ threefold $X$, and, since the type IIA string in
ten-dimensions is dual to M-theory on a circle, we complete the chain
of dualities to arrive at M-theory compactified on $X\times
T^2$.\footnote{Since there heterotic string is actually compactified on
  $T^3$, rather than $T^2$, there are more possibilities for choices
  of Wilson lines as explained in \cite{many:T3}. We ignore this fact.}

Thus we may explicitly map between dual pairs of $S_1\times S_2$ and
$X\times T^2$. The number of families of \CY\ threefolds up to
birational equivalence is known to be at least in the
thousands. What's more, most (although not all) of these \CY's have
moduli spaces which are connected into a big web. Thus, the M-theory
compactifications can be followed through transitions into thousands
of possibilities. The same must therefore be true on the
$\KIII\times\KIII$ side of the duality. But there is only one class of
$\KIII\times\KIII$! This mismatch is solved by allowing for $G$-fluxes
and M2-branes\footnote{To simplify discussion, we will often refer to
``a choice of M2-branes and $G$-fluxes'' simply as a ``choice of
$G$-fluxes''.} on the $\KIII\times\KIII$ side. It turns out that there
are many, many ways of turning on $G$-flux for M-theory on
$\KIII\times\KIII$ while preserving the $N=4$ supersymmetry
\cite{SVW:8}.

Before turning to many examples of these possibilities, let us review
a bit more about the moduli space of these M-theory compactifications.
The $R$-symmetry of $N=4$ in three dimensions is
$\SO(4)\cong\Sp(1)\times\Sp(1)$. We therefore expect the moduli space
to be (locally) of the form $\cM_1\times\cM_2$, where $\cM_1$ and
$\cM_2$ are quaternionic K\"ahler manifolds.

This structure arises in an obvious way for M-theory compactified on
$S_1\times S_2$ --- we associate $\cM_1$ to the moduli space of $S_1$
and $\cM_2$ to the moduli space of $\cM_2$. In the case of M-theory on
$X\times T^2$, we argue as follows. The type IIA string on a \CY\
threefold has a moduli space of the form $\cM_V\times\cM_H$, where
$\cM_V$ is the special K\"ahler moduli space of vector multiplets and
$\cM_H$ is the quaternionic K\"ahler moduli space of
hypermultiplets. We refer to \cite{me:tasi99} and references therein
for more details. Upon compactification of this four-dimensional
theory on a circle to three dimensions, $\cM_V$ becomes
``quaternionified'' and $\cM_H$ is unchanged. Exchanging these factors
amounts to mirror symmetry on $X$. Thus, the mirror symmetry of a \CY\
threefold can be understood, via the above duality as an exchange of
two K3 surfaces.

\subsection{No $G$-flux}  \label{s:G=0}

We first review the case of M-theory on $S_1\times
S_2$, where both $S_1$ and $S_2$ are smooth and we
have $G=0$. The standard tadpole cancellation rule states that
\cite{BB:8,SVW:8,DM:8d}
\begin{equation}
  \nMM + \ff12 G^2 = \frac\chi{24} = 24,  \label{eq:tad}
\end{equation}
where $\nMM$ is the number of M2-branes, i.e., points, on $S_1\times
S_2$. Thus, in the initial case of interest, $\nMM=24$.

We would like to find a \CY\ threefold $X_0$, such that this
compactification of M-theory is equivalent to M-theory on $X_0\times
T^2$. This may be done using the F-theory picture of
\cite{MV:F,MV:F2}.

First, M-theory on $S_1$ is dual to the heterotic string 
(either $E_8\times E_8$ or $\spnh$) on $T^3$ \cite{W:dyn}. If
$S_1$ is smooth, then the resulting gauge group in seven dimensions is
$\GU(1)^{22}$ and we have a moduli space
\begin{equation}
  \cM_{1,7\textrm{-dim}} = \GO(\Gamma_{3,19})\backslash
    O(3,19)/(O(3)\times O(19))\times\R_+,  \label{eq:M7}
\end{equation}
where $H^2(S^1)\cong\Gamma_{3,19}$, the even self-dual lattice of
signature (3,19). Since we will obtain several moduli spaces of the
above form, let us use the shorthand notation $\Gr(\Lambda)$ for the
Grassmannian of maximal space-like planes in the space spanned by the
lattice $\Lambda$, divided by the automorphisms of $\Lambda$. That is,
we denote the above moduli space by $\Gr(\Gamma_{3,19})\times \R_+$.

Now we further compactify this seven-dimensional heterotic theory on
$S_2$. This compactification requires a choice of a bundle $E\to
S_2$. Since we chose $S_1$ generically, there are no non-abelian
groups to be used in the construction of $E$. Furthermore, as we will
see in section \ref{ss:K3sm}, we may not use nontrivial line bundles
for $E$ either. So $E$ must be a completely trivial bundle. That said,
in order to satisfy anomaly cancellation, this bundle should have
$c_2=24$. This apparent contradiction may be evaded by using
point-like instantons \cite{W:small-i,SW:6d} (perhaps more properly
thought of as ideal sheaves \cite{AD:tang}). We therefore require that
the heterotic compactification on $S_2$ has 24 point-like instantons.

The moduli space of $N=(4,4)$ conformal field theories on $S_2$ is
given by $\Gr(\Gamma_{4,20})$ \cite{Sei:K3,AM:K3p,me:lK3}. Here
$\Gamma_{4,20}$ is given by the total cohomology $H^0\oplus H^2\oplus
H^4$ of $S_2$. However, the heterotic string on a K3 surface is not an
$N=(4,4)$ theory. Instanton effects, from the heterotic 5-brane
wrapping K3$\times T^2$, will deform the metric of this moduli
space. Let us, for the time being, ignore these instanton
effects. This would make the moduli space
\begin{equation}
  \cM_2 = \Gr(\Gamma_{4,20}) \ltimes \Sym^{24}(S_2), \label{eq:m2}
\end{equation}
where the $\Sym^{24}(S_2)$ factor arises from the location of the 24
identical instantons on the K3 surface $S_2$. The symbol ``$\ltimes$''
is used to denote a warped product --- the shape of $S_2$ depends on
the moduli in the first factor.

The instantons will, of course, warp the moduli space
(\ref{eq:m2}). We only expect this form of the moduli space to be
accurate in a neighbourhood where the volume of $S_2$ is large.

After compactifying on $S_2$, we now have a theory with gauge group
$\GU(1)^{22}$. In three dimensions, a vector field my be dualized into a
periodic scalar field. Thus, we acquire 22 more moduli. This enhances
the moduli space (\ref{eq:M7}) to
\begin{equation}
  \Gr(\Gamma_{3,19}) \ltimes \GU(1)^{22} \rtimes \R_+.  \label{eq:para}
\end{equation}
Up to some discrete identifications, this is exactly the decomposition
of $\Gr(\Gamma_{4,20})$ used in studying the moduli space of strings
on K3 surfaces (see \cite{me:lK3} for example). Usually the
$\GU(1)^{22}$ factor represents the $B$-field degree of freedom. Here
the same r\^ole is played by the scalars dual to the $\GU(1)^{22}$
gauge group.

There are still more moduli in the three-dimensional field theory due
to the 24 point-like instantons. If these instantons are of the $E_8$
variety, they produce massless tensor fields in six dimensions
\cite{SW:6d}. This supermultiplet contains one real scalar, plus upon
compactification on $T^3$, we obtain $b_1(T^3)=3$ more scalars. That
is, there are 4 real moduli per point-like instanton. Similarly, such
an instanton of the $\spnh$ persuasion produces a $\Sp(1)$ gauge
symmetry in six dimensions \cite{W:small-i}. Wilson lines for this on
$T^3$ produce 3 moduli, and dualizing the resulting $\GU(1)$ in three
dimensions produces a fourth. Thus, again, each point-like instanton
yields 4 real moduli. 

Finally, it is clear that exchanging $S_1$ and $S_2$ in the
compactification of M-theory on $S_1\times S_2$ should be a
symmetry. This must make the moduli space of the compactification
associated to $S_1$, i.e., $\cM_1$, isomorphic to $\cM_2$ given in
(\ref{eq:m2}). The full moduli space must therefore be of the form
\begin{equation}
\begin{split}
\cM_{G=0} &= (\cM_1\times \cM_2)/Z_2\\
  &= \left(\Gr(\Gamma_{4,20}) \times \Gr(\Gamma_{4,20}) \ltimes 
      \Sym^{24}(S_1\times S_2)\right)/\Z_2. \label{eq:MG0}
\end{split}
\end{equation}
Note that we see clearly the moduli space $\Sym^{24}(S_1\times S_2)$
of the 24 M2-branes. Having said that, let us emphasize again that
this moduli space will be deformed by instantons. In the language of
M-theory, these are M5-branes wrapped around $S_1$ times a 2-sphere in
$S_2$, or $S_2$ times a 2-sphere in $S_1$.

Now we turn to the question of finding a \CY\ threefold $X_{G=0}$ such
that the above compactification is dual to M-theory on $X_{G=0}\times
T^2$. The answer to this question has been known for some time
\cite{MV:F}. In order to be fairly explicit, we will use the language
of hypersurfaces in toric varieties following \cite{Bat:m} to describe
our \CY\ threefolds. Here, a \CY\ is described in terms of a reflexive
lattice polytope in some lattice $\mathbf{N}$. This polytope describes
a toric 4-fold. The \CY\ is then realized as a smooth representative
of the anticanonical divisor of this toric variety. It is by no means
true that all \CY\ manifolds can be realized as a hypersurface in a
toric 4-fold but, fortunately, all the manifolds we require in this
paper are of this type.

It was established in \cite{MV:F,MV:F2,FMW:F,AM:po,me:hyp} that the
$E_8\times E_8$ heterotic string on a smooth $\KIII\times T^2$ with a
generic $E_8\times E_8$ bundle with $c_2=(12+n,12-n)$ is dual to a
type IIA string compactified on a \CY\ threefold $X$ specified by a
generic elliptic fibration, with a section, over the Hirzebruch
surface $\HS n$.

An elliptic fibration with a section over $\HS2$ is provided by the
resolution of the hypersurface
\begin{equation}
  x_0^2 + x_1^3 + x_2^{12} + x_3^{24} + x_4^{24}=0, \label{eq:F2}
\end{equation}
in the weighted projective space $\P^4_{\{12,8,2,1,1\}}$ as studied in
\cite{VW:pairs}. This corresponds to the lattice polytope with
vertices
\begin{align} \label{eq:X00}
&(1,0,0,0)&
&(0,1,0,0)&
&(0,0,1,0)&\\  
&(0,0,0,1)&
&(-12,-8,-2,-1).&\notag  
\end{align}
Note that (\ref{eq:F2}) represents the special ``Fermat form'' of the
hypersurface. This may be deformed to include many more monomials. In
lattice language, the terms written in (\ref{eq:F2}) represent the
vertices of the Newton polytope of all possible monomials which appear
in the defining equation for the hypersurface. The Newton polytope is
the polar polytope of that given by (\ref{eq:X00}).

F-theory compactified on the \CY\ given by (\ref{eq:F2}) yields no
gauge symmetry in six dimensions. The bundle structure group has
broken the entire $E_8\times E_8$. We want the opposite extreme where
the bundle is given by point-like instantons and the $E_8\times E_8$
is unbroken. This may be achieved \cite{MV:F} by deforming the above
\CY\ such that it acquires two curves of $E_8$ singularities. The
following form of the hypersurface achieves this:
\begin{equation}
  x_0^2 + x_1^3 + x_2^7x_3^{10}+x_2^7x_4^{10} + 
x_2^5x_3^{14} + x_2^5x_4^{14}=0. \label{eq:F2a}
\end{equation}
This singular \CY\ threefold may be blown-up (corresponding to giving
vacuum expectation values to the tensor moduli). This extremal
transition can be described in terms of lattice polytopes using the
ideas of \cite{ACJM:srch,CGGK:srch}. The monomials in (\ref{eq:F2a})
represent the vertices of a new smaller Newton polytope. The polar of
this Newton polytope provides the new reflexive polytope for the new
manifold. This latter polytope has vertices
\begin{align} \label{eq:XG0}
&(1,0,0,0)&
&(0,1,0,0)&
&(0,0,0,1)&\\
&(-12,-8,-2,-1)&
&(15,10,6,0)&
&(-21,-14,-6,0)\notag  
\end{align}
Let us denote this new \CY\ manifold $X_{G=0}$. Since it corresponds
to 24 point-like instantons, it must be true that the above
compactification of M-theory on $S_1\times S_2$ is dual to the type
IIA string compactified on $X_{G=0}\times S^1$, i.e., M-theory
compactified on $X_{G=0}\times T^2$.

Lest the reader doubt the construction, let us check that the Hodge
numbers agree. M-theory compactified on $X\times T^2$ has a moduli
space generically of the form $\cM_1\times \cM_2$, where $\cM_1$ and
$\cM_2$ are quaternionic K\"ahler manifolds of quaternionic dimension
$h^{1,1}(X)+1$ and $h^{2,1}(X)+1$ respectively. Thus, in order to
match (\ref{eq:MG0}) we require
$h^{1,1}(X_{G=0})=h^{2,1}(X_{G=0})=43$. It is a simple matter
\cite{Bat:m,KS:palp} to check that this is so.

Clearly exchanging $S_1$ and $S_2$ swaps the two factors of the
moduli space and thus corresponds to mirror symmetry. It must
therefore be that $X_{G=0}$ is self-mirror. Since the polytope
(\ref{eq:XG0}) is not self-polar, this fact is not manifest from the
toric description.

\subsection{A singular K3$\times$K3 with $G$-flux}  \label{ss:K3sing}

A \CY\ threefold can undergo extremal transitions (e.g. conifold
transforms) changing the topology but without destroying any
finiteness of a string compactification \cite{GMS:con}. Indeed the
resulting connected ``web'' of components of the moduli space seems to
contain a very large number of the possible
\CY\ threefolds. This must mean that the M-theory compactified on
$S_1\times S_2$ needs to undergo similar transformations.

The simplest transformation corresponds to giving the point-like
instantons of the previous section nonzero size to yield a smooth
bundle for the heterotic string. This picture is again fairly
well-known but we will review the ideas once more to fix notions for
later sections.

In order that the heterotic string on $S_2$ has a nonabelian
structure group, we must have some nonabelian gauge group before
compactification. Thus, M-theory on $S_1$ gives some nonabelian
factors to the gauge group which, in turn, implies that $S_1$ acquires
at least one ADE-like singularity.

\begin{figure}
$$
\setlength{\unitlength}{0.0005in}%
\begin{picture}(7525,5295)(589,-4873)
\thinlines
\put(950,-3192){\circle*{50}}
\put(1085,-3312){\circle*{50}}
\put(950,-3597){\circle*{50}}
\put(1415,-3372){\circle*{50}}
\put(1610,-3237){\circle*{50}}
\put(1715,-3477){\circle*{50}}
\put(1985,-3267){\circle*{50}}
\put(1895,-3057){\circle*{50}}
\put(1715,-3777){\circle*{50}}
\put(1595,-4017){\circle*{50}}
\put(1250,-4317){\circle*{50}}
\put(1850,-4272){\circle*{50}}
\put(2390,-3387){\circle*{50}}
\put(2195,-3147){\circle*{50}}
\put(1066, 14){\circle*{50}}
\put(931,-121){\circle*{50}}
\put(1066,-241){\circle*{50}}
\put(931,-526){\circle*{50}}
\put(1258,-504){\circle*{50}}
\put(1396,-301){\circle*{50}}
\put(1591,-166){\circle*{50}}
\put(1696,-406){\circle*{50}}
\put(1921,-586){\circle*{50}}
\put(2026,-361){\circle*{50}}
\put(1966,-196){\circle*{50}}
\put(1876, 14){\circle*{50}}
\put(1696,-706){\circle*{50}}
\put(1336,-781){\circle*{50}}
\put(1576,-946){\circle*{50}}
\put(1231,-1246){\circle*{50}}
\put(1831,-1201){\circle*{50}}
\put(2161,-931){\circle*{50}}
\put(2311,-1321){\circle*{50}}
\put(1846,-1471){\circle*{50}}
\put(2371,-616){\circle*{50}}
\put(2371,-316){\circle*{50}}
\put(2176,-76){\circle*{50}}
\put(871, 74){\circle*{50}}
\put(1096,-3946){\circle*{50}}
\put(1321,-3646){\circle*{50}}
\put(1276,-3061){\circle*{50}}
\put(2101,-3511){\circle*{50}}
\put(1951,-3811){\circle*{50}}
\put(2146,-4171){\circle*{50}}
\put(2551,-4336){\circle*{50}}
\put(1576,-4486){\circle*{50}}
\put(2101,-4486){\circle*{50}}
\put(2296,-3796){\circle*{50}}
\put(601,-1861){\framebox(2100,2100){}}
\put(601,-4861){\framebox(2100,2100){}}
\put(751,314){\makebox(0,0)[lb]{\smash{$S_1$}}}
\put(751,-2686){\makebox(0,0)[lb]{\smash{$S_2$}}}
\put(1576,-2386){\makebox(0,0)[lb]{\smash{$\times$}}}
\put(6351,-3192){\circle*{50}}
\put(6486,-3312){\circle*{50}}
\put(6351,-3597){\circle*{50}}
\put(6816,-3372){\circle*{50}}
\put(7011,-3237){\circle*{50}}
\put(7116,-3477){\circle*{50}}
\put(7386,-3267){\circle*{50}}
\put(7296,-3057){\circle*{50}}
\put(7116,-3777){\circle*{50}}
\put(6996,-4017){\circle*{50}}
\put(6651,-4317){\circle*{50}}
\put(7791,-3387){\circle*{50}}
\put(7596,-3147){\circle*{50}}
\put(6467, 14){\circle*{50}}
\put(6332,-121){\circle*{50}}
\put(6467,-241){\circle*{50}}
\put(6332,-526){\circle*{50}}
\put(6659,-504){\circle*{50}}
\put(6797,-301){\circle*{50}}
\put(6992,-166){\circle*{50}}
\put(7097,-406){\circle*{50}}
\put(7322,-586){\circle*{50}}
\put(7427,-361){\circle*{50}}
\put(7367,-196){\circle*{50}}
\put(7277, 14){\circle*{50}}
\put(7097,-706){\circle*{50}}
\put(6737,-781){\circle*{50}}
\put(6977,-946){\circle*{50}}
\put(6632,-1246){\circle*{50}}
\put(7772,-616){\circle*{50}}
\put(7772,-316){\circle*{50}}
\put(7577,-76){\circle*{50}}
\put(6272, 74){\circle*{50}}
\put(6497,-3946){\circle*{50}}
\put(6722,-3646){\circle*{50}}
\put(6677,-3061){\circle*{50}}
\put(7502,-3511){\circle*{50}}
\put(7352,-3811){\circle*{50}}
\put(6977,-4486){\circle*{50}}
\put(7697,-3796){\circle*{50}}
\put(7501,-1261){\circle{80}}
\put(3151,-2311){\vector( 1, 0){2250}}
\put(6002,-1861){\framebox(2100,2100){}}
\put(6002,-4861){\framebox(2100,2100){}}
\put(7201,-4036){\line( 0,-1){450}}
\put(7276,-4036){\line( 0,-1){450}}
\put(7351,-4036){\line( 0,-1){450}}
\put(7426,-4036){\line( 0,-1){450}}
\put(7501,-4036){\line( 0,-1){450}}
\put(7576,-4036){\line( 0,-1){450}}
\put(7651,-4036){\line( 0,-1){450}}
\put(7726,-4036){\line( 0,-1){450}}
\put(7801,-4036){\line( 0,-1){450}}
\put(6152,314){\makebox(0,0)[lb]{\smash{$S_1$}}}
\put(6152,-2686){\makebox(0,0)[lb]{\smash{$S_2$}}}
\put(6977,-2386){\makebox(0,0)[lb]{\smash{$\times$}}}
\put(7026,-1486){\makebox(0,0)[lb]{\smash{
\scriptsize$\C^2/\Z_2$-singularity}}}
\put(8200,-4336){\makebox(0,0)[lb]{\smash{$\SU(2)$-bundle}}}
\end{picture}
$$
\caption{A simple transition to a finite-sized instanton.}
\label{fig:t1}
\end{figure}
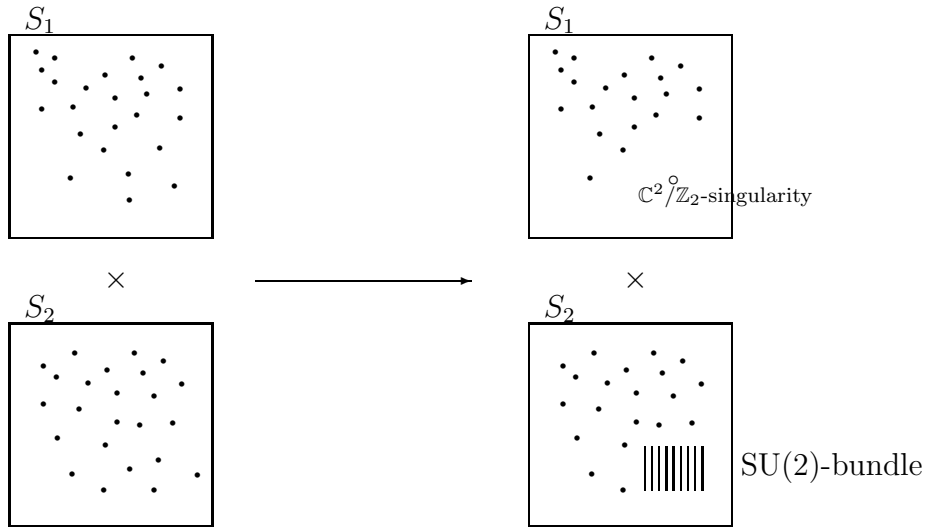

Let $H$ denote a subgroup of $\Sl(2,\Z)$ and let $\cH$ denote the
corresponding Lie group. If $S_1$ acquires a singularity of the form
$\C^2/H$, then the heterotic bundle $E$ on $S_2$ may have structure
group $\cH$. Since $c_2(E)>0$, some of the point-like instantons must
have been eaten up by this bundle. The point-like instantons that
remain uneaten will still correspond to M2-branes in the M-theory
picture of the compactification. These M2-branes are not associated
with the transition and therefore are still free to wander about
$S_1\times S_2$. In order for this description to be consistent,
exactly $c_2(E)$ of the corresponding points on $S_1$ must have gone
into the singularity $\C^2/H$ on $S_1$. We show an example of such a
transition in figure \ref{fig:t1} for an $\SU(2)$-bundle with $c_2=4$.

The interpretation of this transition in terms of $G$-flux is
straight-forward. The supergravity analysis of
supersymmetry-preserving $G$-flux on a four-fold dictates that $G$
must be of type $(2,2)$ and primitive \cite{BB:8}. Following
\cite{DRS:M-G}, one achieves this by putting
\begin{equation}
  G = \sum_{\alpha} \omega_1^{(\alpha)}\wedge \omega_2^{(\alpha)},
     \label{eq:G4}
\end{equation}
where $\omega_j^{(\alpha)}$ is a primitive $(1,1)$-form on $S_i$.

The classical moduli space of Ricci-flat metrics on a K3 surface $S$
is given by $\Gr(\Gamma_{3,19})$ --- the Grassmannian of space-like
3-planes $\Sigma$ in $H^2(S,\R)$. The 3-plane $\Sigma$ is spanned by
the real and imaginary parts of the holomorphic 2-form on $S$ together
with the K\"ahler form $J$ (see \cite{me:lK3} for more details). The
statement that a 2-form $\omega$ is primitive and of type (1,1) is
therefore equivalent to the statement that $\omega$ is perpendicular
to $\Sigma$ in $H^2(S,\R)$.

A K3 surface is hyperk\"ahler. Thus, for a fixed choice of Ricci-flat
metric, there is a whole $S^2$ of compatible complex structures. The
statement that $\omega$ is a primitive (1,1)-form is unaffected by
this choice of complex structure. This means that the hyperk\"ahler
structure is unaffected by the presence of flux. Since the
supersymmetry generators can be constructed from the complex
structures, this means that the $G$-flux given in (\ref{eq:G4}) breaks
none of the extended supersymmetry --- we still have $N=4$ in three
dimensions. Conversely, the only choice of $G$-flux which preserves
$N=4$ supersymmetry must be of the form (\ref{eq:G4}). An example of a
flux which breaks supersymmetry to $N=2$ was given in \cite{DRS:M-G}.

Since we are looking for transitions given by extremal transitions of
\CY\ manifolds with no flux involved, we are not breaking any
supersymmetry and so we need the $G$-flux to be of the form
(\ref{eq:G4}) on $S_1\times S_2$.

Since $G$ is integral, the 2-forms $\omega_j^{(\alpha)}$ can be chosen to be
integral. If $\omega$ is a primitive integral
$(1,1)$-form on a K3 surface $S$, then $\omega^2$ must be a negative
even integer. The minimal flux is therefore given by $\omega^2=-2$. In
this case let $L$ be a line bundle with $c_1(L)=\omega$. The
Riemann-Roch theorem yields
\begin{equation}
\begin{split}
\chi(L) &= h^0(L) - h^1(L) + h^2(L)\\
  &= h^0(L) - h^1(L) + h^0(L^{-1})\\
  &= \int_S \ch(L)\td(S)\\
  &= \ff12\omega^2 + 2\\
  &= 1.
\end{split}
\end{equation}
Thus, either $L$ or $L^{-1}$ has nontrivial sections. The zero-set of
such sections will be an algebraic curve in $S$ Poincar\'e dual to
$\pm\omega$. But the fact that $\omega$ is primitive means that this
algebraic curve has area $J.\omega=0$. This implies that $S$ {\em
must\/} be singular.

The M-theory interpretation of the case in hand is therefore that we
have a $G$-flux given, in part, by a primitive $(1,1)$-form $\omega_1$
on $S_1$ of length-squared $-2$. The rest of the $G$-flux is given by
the curvature of the bundle $E$ on $S_2$. This part of the $G$-flux is
Lie algebra valued and has therefore become nonabelian in some
sense. This is not unreasonable since the connection on the nontrivial
bundle $E$ arises from the M-theory 3-form potential compactified on a
vanishing 2-sphere in $S_1$.

The statement that $\omega_2$ is primitive of type (1,1) should
therefore be naturally stated in terms of $c_1(E)$ being primitive.
Note that since $c_1(E)=0$ for a semi-simple structure group, this
condition is trivial. Therefore, the $G$-flux restricts the moduli of
$S_1$ forcing it to be singular, whereas $S_2$ is still allowed to be
{\em any\/} K3 surface. Indeed, the nonabelian $G$-flux corresponding
to the curvature of the bundle on $S_2$ gives {\em more\/} moduli
associated to $S_2$ as one would expect from an extremal transition.

Clearly in this case of $G$-flux, the $\ff12 G^2$ term in the tadpole
condition (\ref{eq:tad}) is replaced by $c_2(E)$. It is also apparent
that we may mix the r\^oles of $S_1$ and $S_2$. That is, both $S_1$
and $S_2$ may be forced to be singular by $G$-flux and both surfaces
may be endowed with nontrivial gauge bundles with semi-simple
structure groups.

Finally let us note in this section that this picture of M-theory on
$S_1\times S_2$ gives a nice explanation of an observation in
\cite{PR:hetmir}. As we have already argued, mirror symmetry of the
\CY\ threefold $X$ corresponds to exchanging the K3 surfaces $S_1$ and
$S_2$. As seen in figure \ref{fig:t1}, on $S_1$ a number, $n$, of
point-like instantons have merged into a quotient singularity
$\C^2/H$, whereas on $S_2$ we have obtained a smooth bundle with
structure group $\cH$ and with $c_2=n$. Thus mirror symmetry exchanges
$n$-point like instantons embedded in a quotient singularity $\C^2/H$,
with a smooth bundle with structure group $\cH$ and with
$c_2=n$. Also, a point-like instanton at a smooth point is
exchanged with another point-like instanton at a smooth point. This is
exactly the phenomenon experimentally observed in \cite{PR:hetmir}.

\subsection{Smooth K3 surfaces and $\GU(1)$-bundles} \label{ss:K3sm}

The picture of the last section gave a fairly nice interpretation of
$G$-flux obstructing moduli, in that it forced $S_1$ to be singular. It
would be more satisfying, though, to give an example of moduli
obstruction where the K3 surfaces were smooth.

To do this, let us consider a specific \CY\ threefold and prove in
detail an interpretation of F-theory compactified on this threefold
conjectured in \cite{AFIU:U1}.

The type IIA string compactified on any \CY\ threefold $X$ is, in some
sense, always dual to a heterotic string compactified on $\KIII\times
T^2$. Having said that, the heterotic string cannot be weakly-coupled
at any point in the moduli space unless $X$ is a K3 fibration
\cite{KLM:K3f,AL:ubiq}. Furthermore, the $T^2$ can only be
decompactified to produce a six-dimensional, F-theory,
compactification if each K3 fibre is itself an elliptic fibration with
a section. Thus we will restrict ourselves to the case where $X$ is an
elliptic fibration with a section.

In any such case, one can very systematically
\cite{MV:F,MV:F2,FMW:F,AM:po,me:hyp} establish precisely to which
$E_8\times E_8$ heterotic string compactification this F-theory
compactification is dual.\footnote{The techniques are not quite so
well-developed for the $\spnh$ case at present.}

The type IIA string compactified on $X$ will generically produce
$h^{1,1}(X)$ vector multiplets in four dimensions. These have two
sources from a six-dimensional theory compactified on $T^2$. They may
originate either from six-dimensional vector or tensor
multiplets. Which is the case was explained in \cite{MV:F,MV:F2}. The
result is as follows. Consider an F-theory compactification on $X$, where
$X$ is an elliptic fibration $\pi:X\to B$. The associated spectral
sequence for cohomology implies that $H^2(X,\Z)$ has three sources
which have the effect:
\begin{enumerate}
\item $H^2(B,\Z)$ from the base which yield tensor multiplets (except
  for two),
\item $H^0(B,R^2\pi_*\Z)$ from the fibres which yield vector
  multiplets (except for one), and
\item $H^1(B,R^1\pi_*\Z)$ which also yield vector multiplets.
\end{enumerate}

The $H^0(B,R^2\pi_*\Z)$ contribution, beyond one coming from the
generic fibre, comes from singular fibres with more than one
component. These produce nonabelian gauge symmetries in six
dimensions. The $H^1(B,R^1\pi_*\Z)$ contribution does not produce
nonabelian symmetries and therefore is associated to $\GU(1)$ factors
in the six-dimensional gauge group.

A key notion in the study of elliptic fibrations is the {\em
Mordell-Weil\/} group $\Phi$. This is the group of sections (viewing
an elliptic curve as the group $\GU(1)\times\GU(1)$). Being a finitely
generated abelian group, $\Phi$ will have some finite torsion part and
a free part of finite rank, $\rank(\Phi)$. For an elliptic fibration,
the part of $H^1(B,R^1\pi_*\Z)$ that is of type $(1,1)$ will provide
the free part of $\Phi$ (see \cite{CZ:gag} for example). Thus, {\em the
number of $\GU(1)$ factors of the six-dimensional gauge group in
F-theory is given by the rank of the Mordell--Weil group\/} as stated in
\cite{MV:F2}.

One can also show \cite{AM:frac} that the torsion part of $\Phi$ is
associated to the gauge group being not simply-connected. Thus $\pi_1$
of the six-dimensional gauge group is isomorphic to $\Phi$.

Armed with the necessary facts, let us consider the \CY\ given by the
hypersurface
\begin{equation}
  x_0^2 + x_1^4 + x_2^8 + x_3^{16} + x_4^{16},  \label{eq:X1f}
\end{equation}
in $\P^4_{\{8,4,2,1,1\}}$. If we fix the values of $[x_2,x_3,x_4]$ we
are left with the elliptic curve of degree 4 in
$\P^2_{\{2,1,1\}}$. Thus, this threefold contains a ``net'' (i.e., a
two-dimensional family) of elliptic curves. This does not mean that
our threefold is elliptic however.

The space (\ref{eq:X1f}) contains two singularities of the form
$\C^3/\Z_4$ at the two points $[x_0,x_1,x_2,\allowbreak x_3,x_4]=[\pm
i,1,0,0,0]$. Every elliptic curve in our net passes through these two
points. Blowing up these two points (together with a curve connecting
them) will resolve the hypersurface. It will also make each elliptic
curve in the net disjoint and thus yield an elliptic fibration. Let us
denote this elliptic threefold $X_1$.

The two exceptional divisors of this resolution each provide a section
of the elliptic fibration. Thus $\Phi$ contains at least two elements.
If we use a {\em generic\/} equation for the hypersurface (rather
than the Fermat form (\ref{eq:X1f})) then all the fibres
have only one component and thus $H^0(B,R^2\pi_*\Z)$ is rank one. 
The base of the fibration is $\HS2$ and thus $H^2(B,\Z)$ is rank two.
It is easy to show that $h^{1,1}(X_1)=4$ and thus $H^1(B,R^1\pi_*\Z)$
is of rank one. That is, the group of sections is infinite. We can
view one of the exceptional divisors as the ``zero section'' $\sigma_0$. The
other exceptional divisor $\sigma_1$ generates $\Phi\cong\Z$.

For special choices of complex structure, such as the one
corresponding to the Fermat form (\ref{eq:X1f}), $\sigma_1$ actually
corresponds to a torsion section and the Mordell--Weil group is only
$\Z_2$. One also acquires fibres with more than one component. The
standard F-theory interpretation of this would then have at least an
$\SO(3)$ gauge group. To avoid this subtlety for now, we will assume
that $X_1$ has a generic complex structure but we will visit
this issue again at the end of section \ref{ss:btof}.

Thus, F-theory compactified on $X_1$ yields a gauge group $\GU(1)$ in
six-dimensions as conjectured in \cite{AFIU:U1}.\footnote{This fact is
disputed in \cite{BKMT:IIBv} where it is asserted that $\GU(1)$ gauge
groups arise from the monodromy group of the elliptic fibration lying
in a proper subgroup of $\Sl(2,\Z)$. We believe that the authors of
\cite{BKMT:IIBv} were misled by the fact that, as noted above, for
special values of complex structure, a free section may indeed become
a torsion section. Torsion sections are associated with the
restriction of the monodromy group but free sections are not (see
\cite{Pers:RES} for examples).}
Indeed, every one of the conjectured models in \cite{AFIU:U1} can be
analyzed similarly. For example, if the elliptic curves are cubic
equations in $\P^2$ one will have to blow up three base points of the
net of curves producing a Mordell-Weil group isomorphic to
$\Z^{\oplus2}$ and thus a gauge group of $\GU(1)^2$.

We can now follow the arguments of \cite{GSW:a6,Wit:O(32),AFIU:U1} to
see how this $\GU(1)$ gauge group can arise in the language of
the $E_8\times E_8$ heterotic string. The effective action of a
heterotic compactification contains the term $H^2$, where
\begin{equation}
  H = dB + \omega_L - \omega_Y.
\end{equation}
A generic $\GU(1)$-bundle with a generic $c_1$ will produce a vacuum
expectation value for the bundle curvature $\langle F\rangle$. This,
in turn, will produce a term proportional to $AdB$ in the effective
action. By analogy with the Higgs' mechanism, a coordinate change in
these fields will have the effect of making the
$\GU(1)$ photon massive while removing a zero mode of the
$B$-field. That is, the $\GU(1)$ does not actually appear as a gauge
symmetry. In order to circumvent this effect we take a $\GU(1)$-bundle
with nonzero $c_1$ and embed it identically into both $E_8$ factors in
the heterotic string. The diagonal combination of these $\GU(1)$'s
will gain a mass by the above mechanism, but the anti-diagonal
combination will not, thus leaving a single $\GU(1)$ gauge group as a
low-energy symmetry.

This idea is confirmed if we use the stable degeneration method of
\cite{FMW:F,AM:po,me:hyp} to find precisely the heterotic dual of
F-theory on $X_1$. One takes a limit in the complex structure of $X_1$
which is dual to taking the large $T^2$ limit of the heterotic string
on $\KIII\times T^2$. $X_1$ then degenerates into a pair of elliptic
threefolds $X_1^{(1)}$ and $X_1^{(2)}$ which intersect along a K3
surface $S_H$. The latter K3 surface is identified as the K3 surface
on which the dual heterotic string is compactified.

In order to find the bundle on $S_H$ for the heterotic string, one
notices that $X_1^{(1)}$ and $X_1^{(2)}$ are fibrations over $\P^1$
with fibre given by a rational elliptic surface. The generators of the
Mordell-Weil group of these rational elliptic surfaces intersect $S_H$ along
curves $C^{(1)}$ and $C^{(2)}$. These curves provide ``cameral''
curves that encode the data of $E_8$-bundles. We refer to
\cite{FMW:F,Don:F,me:hyp} for more information.\footnote{The reference
  \cite{BS:K3K3} also used spectral curves to analyze the $\GU(1)$
  bundles but in a different way to the one presented here.}

The key point is that the Mordell--Weil group of $X_1$ has a
non-trivial generator. This generator is preserved by the stable
degeneration and continues to have its effect on the cameral curve. To
be precise, $S_H$ is itself an elliptic K3 surface with a section
$\sigma_1$ generating a free component of its own Mordell--Weil
group. The r\^ole of our global section in the Mordell--Weil groups of
$X_1^{(1)}$ and $X_1^{(2)}$ will mean that both $C^{(1)}$ and
$C^{(2)}$ contain a component given by $\sigma_1$.

A rational curve component to the cameral curve signifies a $\GU(1)$
structure group of a factor of the gauge bundle. Thus we explicitly
see a $\GU(1)$-gauge bundle embedded identically into both $E_8$
factors. We may do even better than this. The gauge bundle can be
constructed as a Fourier--Mukai transform of the cameral cover data
(at least in the case of holomorphic bundles with structure group
$\GU(N)$ which is good enough for our purposes here). Thus produces
the Chern class data for the bundle. We refer to \cite{AD:tang} for
the relevant information. The only fact we require here is that the
Fourier--Mukai transform only acts on a 4-dimensional sublattice of
$H^*(\KIII)$ generated by $H^0$, $H^4$, the 2-form dual to the
elliptic fibre of the K3 surface, and the 2-form dual to the zero
section. Directions in $H^2$ orthogonal to this lattice are unchanged
by the transformation. This allows us to compute $c_1$ of the heterotic
bundle. The result is (at least up to unimportant fibre contributions)
$c_1=\sigma_1-\sigma_0$ --- the difference between the two sections.

For the case of the \CY\ threefold $X_1$, the bundle structure group
is generically\footnote{See section \ref{ss:btof} for an explanation
of the $\Z_2$ quotient.} $((E_7\times \GU(1))/\Z_2)^{\oplus2}$ leaving
an unbroken gauge group of $\GU(1)$. Returning to the language of
M-theory on $S_1\times S_2$, we saw in section \ref{ss:K3sing} that
bundles with a nonabelian structure group corresponded to singular K3
surfaces. We would like to make the K3 surfaces $S_1\times S_2$
smooth. In analogy with section \ref{s:G=0}, we would therefore like
to reduce the structure group of the heterotic bundle as much as
possible and produce point-like instantons.

If we reduce the $E_7$ parts of the structure group completely we
would expect to obtain an $(E_7\times E_7\times\GU(1)/\Z_2)$
six-dimensional gauge group from F-theory. This may be achieved by
deforming the complex structure of $X_1$ to acquire two curves of type
$\textrm{III}^*$ fibres by choosing a hypersurface with equation
\begin{equation}
  x_0^2 + x_1^4 + x_2^5x_3^6 + x_2^5x_4^6 + x_2^3x_3^{10} + x_2^3x_4^{10}.  
      \label{eq:X1E7}
\end{equation}
As in section \ref{s:G=0}, this produces an extremal transition to a
new manifold we call $X_{U(1)}$, which is a hypersurface in a toric
variety associated to a polytope with vertices:
\begin{align} \label{eq:XU1}
&(1,0,0,0)&
&(0,1,0,0)&
&(0,0,0,1)&\\
&(-8,-4,-2,-1)&
&(6,3,4,0)&
&(-10,-5,-4,0).\notag  
\end{align}
It is an easy matter to show that
$h^{1,1}(X_{U(1)})=h^{2,1}(X_{U(1)})=34$ and thus, for M-theory
compactified on $X_{U(1)}\times T^2$,
$\dim\cM_1=\dim\cM_2=35$. Examination of the discriminant locus of the
elliptic fibration (\ref{eq:X1E7}) shows that there are 16 point-like
instantons of the $E_8$ kind. 

We now claim that M-theory compactified on $X_{U(1)}\times T^2$ is
dual to a particular flux-compactification of M-theory compactified on
$S_1\times S_2$. The $G$-flux is given by $\omega_1\wedge\omega_2$,
where $\omega_j=\alpha_j-\beta_j$, and $\alpha_j$ and $\beta_j$ are
dual to two disjoint $(-2)$-curves (i.e., $S^2$'s) in $S_j$. Thus,
$G^2=16$, consistent with the rest of the tadpole cancellation being
given by 16 M2-branes. This model first appeared in
\cite{DRS:M-G}. The 2-forms $\alpha_1$ and $\beta_1$ are associated
with the embedding of the $\GU(1)$ bundles in each $E_8$. The 2-forms
$\alpha_2$ and $\beta_2$ are directly identified with the sections
$\sigma_0$ and $\sigma_1$. We now explain the correspondence in
detail.

Let us first make a general statement about the M-theory
interpretation of moduli obstructions due to $G$-flux. The statement
that $\omega_1$ or $\omega_2$ is (1,1) and primitive can never give a
full description of the obstructions. This is because such a statement
removes real metric moduli in multiples of 3 (2 from the (1,1)
statement and 1 for the primitive statement). The moduli space is
quaternionic K\"ahler and so obstructions should be removed in
multiples of 4. Thus, there should always be some extra effect that
removes moduli beyond the primitive-(1,1) condition. We find this to
be the case in this example.

Recall that earlier our moduli spaces contained factors like
$\Gr(\Lambda_{4,20})$ corresponding a Grassmannian of space-like
4-planes. When explicitly constructing these Grassmannians in the
current context, the space like 4-plane, which we denote $\Pi$, is
naturally described as being spanned by a space-like 3-plane $\Sigma$
(associated to the metric on the K3 surface) and fourth direction. We
use this language below.

Let us first interpret $\omega_1=\alpha_1-\beta_1$. $\alpha_1$ and
$\beta_1$ represent two-spheres in $S_1$ which map to directions in
the $E_8\times E_8$ lattice for the heterotic string. $\alpha_1$ is in
one $E_8$ and $\beta_1$ is in the other. By the rules of $G$-flux
compactifications, $\omega_1=\alpha_1-\beta_1$ is type $(1,1)$ and
$J.\omega=0$. This puts the space-like 3-plane $\Sigma$ perpendicular
to $\omega_1$. The fact that $\omega^2=-4$ means that $S_1$ need not
be singular in contrast to the example in section \ref{ss:K3sing},
since Riemann--Roch no longer implies that $\omega$ corresponds to the
class of an algebraic curve. Since one of the $\GU(1)$ gauge fields
becomes massive via the Higgs' mechanism, we lose one of the
$\GU(1)$'s in (\ref{eq:para}). This means, in the language of the
Grassmannian $\Gr(\Gamma_{4,20})$, the space-like 4-plane $\Pi$ is
perpendicular to $\omega_1$. Thus $\omega_1$ obstructs one
quaternionic deformation of $S_1$.

The form $\omega_2$ will similarly obstruct $S_2$. In fact, we may see
this very explicitly. When we blew-up the hypersurface (\ref{eq:X1f})
in $\P^4_{8,4,2,1,1}$, there was a fixed curve of $\C^2/\Z_2$
singularities along $x_3=x_4=0$. This gets resolved along with the two
$\C^3/\Z_4$ singularities discussed above. When we reach the stable
degeneration, this resolved curve hits $S_H$ twice. That is, $S_H$
contains {\em two\/} $\P^1$'s arising from this resolution. This
forces these two homologically independent curves in $S_H$ to have
equal area. We identify these two curves as dual to $\alpha_2$ and
$\beta_2$ and it immediately follows that $J.(\alpha_2-\beta_2)=0$ as
desired. Obviously $\omega_2=\alpha_2-\beta_2$ is of type $(1,1)$
since it is dual to algebraic cycles. 

\begin{table}
\begin{center}
\begin{tabular}{|p{27mm}|p{30mm}|p{60mm}|}
\hline\hline
Obstruction in Moduli space&$G$-flux language&Heterotic Language\\
\hline\hline
$\Sigma_1\perp\omega_1$&$\omega_1$ is (1,1) and primitive&
There is a $\GU(1)\times\GU(1)$-bundle $L$ 
on $S_2$\\
\hline $\Pi_1\perp\omega_1$&new effect&In addition, one $\GU(1)$
photon acquires a mass\\ 
\hline\hline
$\Sigma_2\perp\omega_2$&$\omega_2$ is (1,1) and primitive& $c_1(L)$ is
primitive.\\ \hline $\Pi_2\perp\omega_2$&new effect&In addition, one
component of $B$-field is eaten\\ \hline\hline
\end{tabular}  
\caption{Obstructing the moduli space in two languages.} \label{tab:obs}
\end{center}
\end{table}

In addition, the Higgs' mechanism ate up one of the $B$-field zero
modes associated with the $\GU(1)$-bundles $c_1$ --- namely
$\omega_2$. Thus $\cM_2$ is also associated with a Grassmannian of
4-planes $\Pi$ perpendicular to $\omega_2$. The remaining 16
point-like instantons are free to move about as M2-branes on
$S_1\times S_2$ in a similar way to section \ref{s:G=0}. The
dimensions of $\cM_1$ and $\cM_2$ should therefore be $16+19$ in
agreement with the Hodge numbers above.

Since this story might be a little hard to follow, we review the key
ideas in table \ref{tab:obs}. The entries labeled ``new effect''
refer to the extra obstructions beyond the primitive-(1,1) condition
required to make the obstructions a multiple of 4.

Finally note that everything is symmetric between $S_1$ and $S_2$ in
this construction. Thus we expect $X_{U(1)}$ to be self-mirror.
To summarize, we have shown in detail that M-theory on 
$X_{U(1)}\times T^2$ is dual to M-theory on $S_1\times S_2$ where $G=
\omega_1\wedge\omega_2$ as above and the moduli space is
\begin{equation}
\begin{split}
\cM_{U(1)} &= (\cM_1\times \cM_2)/\Z_2\\
  &= \left(\Gr(\Gamma_{4,19}) \times \Gr(\Gamma_{4,19}) \ltimes 
      \Sym^{16}(S_1\times S_2)\right)/\Z_2. \label{eq:MU1}
\end{split}
\end{equation}

We may extend this analysis to more than one $\GU(1)$ group. For
example, using another example from \cite{AFIU:U1}, M-theory on a
2-torus times the \CY\ given by the polytope with vertices
\begin{align} \label{eq:XU1U1}
&(1,0,0,0)&
&(0,1,0,0)&
&(0,0,0,1)&\\
&(-4,-4,-2,-1)&
&(2,2,3,0)&
&(-4,-4,-3,0),\notag  
\end{align}
is dual to M-theory on $S_1\times S_2$ with
\begin{equation}
  G = (\alpha_1-\gamma_1)\wedge(\alpha_2-\gamma_2) -
        (\beta_1-\delta_1)\wedge(\beta_2-\delta_2),
\end{equation}
where $\alpha_j,\beta_j,\gamma_j,\delta_j$ are dual to $(-2)$-curves
on $S_j$ with $\alpha_j.\beta_j=\gamma_j.\delta_j=1$ and with all
other intersections zero; and 12 M2-branes. In this case, the
six-dimensional F-theory model would have a gauge group
$\GU(1)\times\GU(1)$. The moduli space (before instantons corrections)
is
\begin{equation}
  \left(\Gr(\Gamma_{4,18}) \times \Gr(\Gamma_{4,18}) \ltimes 
      \Sym^{12}(S_1\times S_2)\right)/\Z_2,
\end{equation}
where the 4-plane $\Pi_j$ is perpendicular to $\alpha_j-\gamma_j$ and
$\beta_j-\delta_j$.

Before closing this section we should point out a subtlety when the
heterotic bundle structure group contains both abelian and
non-abelian factors. Consider the \CY\ $X_1$ given by (\ref{eq:X1f}).
As stated above, the six-dimensional gauge group for F-theory on $X_1$
has a gauge group given only by $\GU(1)$. Denoting the structure group
of the heterotic bundle by $H$, this means that $H$ must centralize
$\GU(1)\times \GU(1)$ in $E_8\times E_8$ (since one $\GU(1)$ was eaten
by the Higgs' mechanism). One might carelessly assert that
$\GU(1)\times E_7$ is a subgroup of $E_8$ but this is not true. A more
careful analysis shows that the correct subgroup is $(\GU(1)\times
E_7)/\Z_2$, where the $\Z_2$ is generated by $-1$ in $\GU(1)$ and
acts as the centre of $E_7$. An obvious choice for $H$ would
therefore be $((\GU(1)\times E_7)/\Z_2)^2$. 

In the case of $X_{U(1)}$ we allowed the heterotic bundle to degenerate
so that its structure group was only $\GU(1)^2$. The first Chern class
of these line bundles then contributed towards the $G$-flux the
equivalent of 8 M2-branes. For $X_1$ we would over-count the
contribution to the tadpole if we added 8 to $c_2$ of the $E_7\times
E_7$ precisely because of this $\Z_2$ quotient. Instead, the effective
$c_1$ of the abelian part of the bundle is halved and thus $c_1^2$
only contributes 2 to the tadpole. The result is, as observed in
\cite{AFIU:U1}, that one requires an $E_7\times E_7$ bundle with $c_2=22$
to produce an anomaly-free theory. Only then do we get the dimension
of the moduli space to agree with the Hodge numbers of $X_1$. 

\subsection{A brane to flux transition} \label{ss:btof}

In section \ref{s:G=0} we had M-theory on $\KIII\times\KIII$ with no
flux and 24 M2-branes; and in section \ref{ss:K3sm} we had a theory
with fluxes and only 16 M2-branes. We may follow extremal transitions
through the dual picture of M-theory on a \CY\ threefold times a
2-torus to see how M2-branes may turn into fluxes. Some of the
analysis we do also appeared in \cite{BS:K3K3} and a similar
transition appeared in \cite{BKMT:IIBv}.

Consider deforming the hypersurface of (\ref{eq:F2}) to the singular
hypersurface $X^\sharp$ given by
\begin{equation}
  x_0^2 + x_1^3 + x_1x_2^5x_3^6 + x_1x_2^5x_4^6 + x_1x_2^3x_3^{10}
   + x_1x_2^3x_4^{10}.   \label{eq:tran}
\end{equation}
The F-theory interpretation of this is as follows. The base $B$ of
this fibration is $\HS2$ (before some of the blow-ups). We have
lines of type $\mathrm{III}^*$ fibres in $B$ along $x_2=0$ and along
$x_2=\infty$. We also have a curve of $\mathrm{III}$ fibres along a
curve which intersects the two $\mathrm{III}^*$ lines 6 times and 10
times respectively. The Mordell--Weil group is $\Z_2$. Thus, the
six-dimensional gauge group is $(E_7\times E_7\times
\SU(2))/\Z_2$. There are 16 point-like instantons coming from the
fibre collisions.

If we follow the stable degeneration, the heterotic string K3 surface,
$S_H$ intersects the line of $\mathrm{III}$ fibres a total of 8 times
and thus has 8 $\C^2/\Z_2$ quotient singularities. Following
\cite{AM:frac} we interpret this as a heterotic string
compactification on $S_H$ in the following way. Let $S_0$ be a smooth K3
surface with a $\Z_2$ symmetry that preserves the holomorphic
2-form. $S_H$ is then constructed as the orbifold $S_0/\Z_2$. Put a
bundle on $S_H$ that is trivial except for monodromy around the 8
orbifold points. This monodromy acts as the $\Z_2$ subgroup of $E_8$
that centralizes $(E_7\times\SU(2))/\Z_2$ in each $E_8$. In addition,
the curve of $\mathrm{III}$ fibres tells us that the $\SU(2)$ group
is diagonally embedded in both $E_8$'s. The only way to express this
in terms of centralizing a group action is to include an {\em
exchange\/} of the two $\SU(2)$ subgroups of $E_8\times E_8$ in the
monodromy of the bundle to obtain the correct unbroken gauge 
group.\footnote{This is a little odd since the apparent structure
group of the bundle is now not in $(E_8\times E_8)\rtimes\Z_2$. This
point deserves to be better-understood.}

As in previous sections, we may resolve the hypersurface $X^\sharp$
given by (\ref{eq:tran}) to form a smooth \CY\ threefold we denote
$X'$. The smooth threefold $X'$ is then given by a hypersurface in
a toric variety associated to a polytope with vertices
\begin{align} \label{eq:X1p}
&(1,0,0,0)&
&(1,2,0,0)&
&(0,0,0,1)&\\
&(-12,-8,-2,-1)&
&(9,6,4,0)&
&(-15,-10,-4,0).\notag  
\end{align}

We now wish to prove that $X'$ is isomorphic to $X_{U(1)}$ of
section \ref{ss:K3sm}. To do this, note that
\begin{equation}
\left(\begin{smallmatrix}
1&0&0&0\\0&1&0&0\\0&0&0&1\\-8&-4&-2&-1&\\6&3&4&0\\-10&-5&-4&0
\end{smallmatrix}\right)
=
\left(\begin{smallmatrix}
1&0&0&0\\1&2&0&0\\-12&-8&-2&-1\\0&0&0&1\\9&6&4&0\\-15&-10&-4&0
\end{smallmatrix}\right)
\left(\begin{smallmatrix}
-2&1&0&0\\1&0&0&0\\0&0&1&0\\16&-12&-2&-1
\end{smallmatrix}\right)
\left(\begin{smallmatrix}
-\frac12&\frac12&0&0\\0&1&0&0\\0&0&1&0\\0&0&0&1
\end{smallmatrix}\right). \label{eq:matrix}
\end{equation}
The matrix on the left hand side of this equation consists of the
lattice vectors (\ref{eq:XU1}) used to make $X_{U(1)}$. The first
factor on the right hand side are the lattice vectors (\ref{eq:X1p})
used to make $X'$. The second factor on the right is an element of
$\Gl(4,\Z)$ and so just a simple change of basis of the lattice
$\mathbf{N}$. 

\begin{figure}
$$
\setlength{\unitlength}{0.000500in}%
\begin{picture}(7900,1200)(439,-1648)
\thinlines
\put(2101,-661){\line( 0,-1){750}}
\put(1501,-1186){\line( 1, 0){1200}}
\put(1801,-1036){\line(-1,-1){600}}
\put(2401,-1036){\line( 1,-1){600}}
\put(1426,-1636){\line(-1, 1){600}}
\put(1051,-1036){\line(-1,-1){600}}
\put(2701,-1636){\line( 1, 1){600}}
\put(3076,-1036){\line( 1,-1){600}}
\put(7726,-1186){\circle*{100}}
\put(7726,-736){\circle*{100}}
\put(8251,-1186){\circle*{100}}
\put(7727,-661){\line( 0,-1){750}}
\put(7127,-1186){\line( 1, 0){1200}}
\put(7427,-1036){\line(-1,-1){600}}
\put(7052,-1636){\line(-1, 1){600}}
\put(6677,-1036){\line(-1,-1){600}}
\put(4126,-811){\vector( 1, 0){1650}}
\put(5776,-1411){\vector(-1, 0){1650}}
\put(4501,-711){\makebox(0,0)[lb]{\smash{$/\Z_2$}}}
\put(4501,-1311){\makebox(0,0)[lb]{\smash{resolve}}}
\end{picture}
$$
\caption{The self-similarity of the type $\mathrm{III}^*$ fibre under a
  $\Z_2$-quotient.} \label{fig:III}
\end{figure}
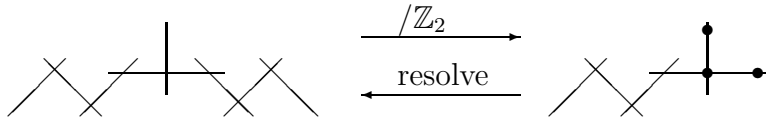
The last factor on the right of (\ref{eq:matrix}) is manifestly not in
$\Gl(4,\Z)$. Thus, the two toric varieties in which $X_{U(1)}$ and
$X'$ are embedded are not manifestly isomorphic. They differ by a
$\Z_2$ refinement of the lattice $\mathbf{N}$. Such a refinement
corresponds to orbifolding by a $\Z_2$ subgroup of the $(\C^*)^4$
torus action. In the case above, it is easy to show that this amounts
to the statement that $X'$ is a resolution of the $\Z_2$ orbifold
of $X_{U(1)}$ given by the generator 
\begin{equation}
g:[x_0,x_1,x_2,x_3,x_4]\to[-x_0,-x_1,x_2,x_3,x_4]. \label{eq:Z2}
\end{equation}
Curiously this orbifolding turns out to have no effect. The action
given by (\ref{eq:Z2}) clearly acts purely on the fibre of the
elliptic fibration leaving the base $\HS2$ untouched. On the smooth
elliptic fibres, the action of $g$ is free and thus we retain the
structure of an elliptic fibration. The only bad fibres in this model
are of type $\mathrm{III}$ or $\mathrm{III}^*$. These bad fibres do
have fixed points but the process of quotienting the fibre and then
blowing up leaves these fibres where they started. For example, we
show what happens to the $\mathrm{III}^*$ fibre in figure
\ref{fig:III}. The $\Z_2$-action is roughly the left-right symmetry in
this figure. The dots on the right-hand side denote fixed points to be
resolved. 

This resolved orbifold also has a section since the zero section of
$X_{U(1)}$ is identified with a torsion section. We have therefore
shown that $X_{U(1)}$ and $X'$ are both elliptic fibrations with a
section with identical fibres in the same configuration over the same
base. It follows \cite{DoGr:} that $X_{U(1)}$ and $X'$ are
birationally equivalent. In this case, since there are generically no
special fibres appearing in codimension 2 over the base, it follows
that $X_{U(1)}$ and $X'$ are isomorphic.

\begin{figure}
$$
\setlength{\unitlength}{0.0004in}%
\begin{picture}(12462,6295)(589,-5873)
\thinlines
\put(950,-3192){\circle*{50}}
\put(1085,-3312){\circle*{50}}
\put(950,-3597){\circle*{50}}
\put(1415,-3372){\circle*{50}}
\put(1610,-3237){\circle*{50}}
\put(1715,-3477){\circle*{50}}
\put(1985,-3267){\circle*{50}}
\put(1895,-3057){\circle*{50}}
\put(1715,-3777){\circle*{50}}
\put(1595,-4017){\circle*{50}}
\put(1250,-4317){\circle*{50}}
\put(1850,-4272){\circle*{50}}
\put(2390,-3387){\circle*{50}}
\put(2195,-3147){\circle*{50}}
\put(1066, 14){\circle*{50}}
\put(931,-121){\circle*{50}}
\put(1066,-241){\circle*{50}}
\put(931,-526){\circle*{50}}
\put(1258,-504){\circle*{50}}
\put(1396,-301){\circle*{50}}
\put(1591,-166){\circle*{50}}
\put(1696,-406){\circle*{50}}
\put(1921,-586){\circle*{50}}
\put(2026,-361){\circle*{50}}
\put(1966,-196){\circle*{50}}
\put(1876, 14){\circle*{50}}
\put(1696,-706){\circle*{50}}
\put(1336,-781){\circle*{50}}
\put(1576,-946){\circle*{50}}
\put(1231,-1246){\circle*{50}}
\put(1831,-1201){\circle*{50}}
\put(2161,-931){\circle*{50}}
\put(2311,-1321){\circle*{50}}
\put(1846,-1471){\circle*{50}}
\put(2371,-616){\circle*{50}}
\put(2371,-316){\circle*{50}}
\put(2176,-76){\circle*{50}}
\put(871, 74){\circle*{50}}
\put(1096,-3946){\circle*{50}}
\put(1321,-3646){\circle*{50}}
\put(1276,-3061){\circle*{50}}
\put(2101,-3511){\circle*{50}}
\put(1951,-3811){\circle*{50}}
\put(2146,-4171){\circle*{50}}
\put(2551,-4336){\circle*{50}}
\put(1576,-4486){\circle*{50}}
\put(2101,-4486){\circle*{50}}
\put(2296,-3796){\circle*{50}}
\put(601,-1861){\framebox(2100,2100){}}
\put(601,-4861){\framebox(2100,2100){}}
\put(751,314){\makebox(0,0)[lb]{\smash{$S_1$}}}
\put(751,-2686){\makebox(0,0)[lb]{\smash{$S_2$}}}
\put(1576,-2386){\makebox(0,0)[lb]{\smash{$\times$}}}
\put(6052,-3192){\circle*{50}}
\put(6187,-3312){\circle*{50}}
\put(6052,-3597){\circle*{50}}
\put(6517,-3372){\circle*{50}}
\put(6712,-3237){\circle*{50}}
\put(6817,-3477){\circle*{50}}
\put(7087,-3267){\circle*{50}}
\put(6997,-3057){\circle*{50}}
\put(7492,-3387){\circle*{50}}
\put(7297,-3147){\circle*{50}}
\put(6168, 14){\circle*{50}}
\put(6033,-121){\circle*{50}}
\put(6168,-241){\circle*{50}}
\put(6033,-526){\circle*{50}}
\put(6360,-504){\circle*{50}}
\put(6498,-301){\circle*{50}}
\put(6693,-166){\circle*{50}}
\put(6798,-406){\circle*{50}}
\put(7068,-196){\circle*{50}}
\put(6978, 14){\circle*{50}}
\put(6798,-706){\circle*{50}}
\put(6438,-781){\circle*{50}}
\put(6678,-946){\circle*{50}}
\put(6333,-1246){\circle*{50}}
\put(7278,-76){\circle*{50}}
\put(5973, 74){\circle*{50}}
\put(6423,-3646){\circle*{50}}
\put(6378,-3061){\circle*{50}}
\put(7203,-3511){\circle*{50}}
\put(7202,-1261){\circle{150}}
\put(7427,-661){\circle{150}}
\put(6302,-3511){\circle*{50}}
\put(6752,-3061){\circle*{50}}
\put(6977,-3661){\circle*{50}}
\put(6002,-3961){\circle{150}}
\put(6302,-3961){\circle{150}}
\put(6602,-3961){\circle{150}}
\put(6902,-3961){\circle{150}}
\put(6902,-4261){\circle{150}}
\put(6602,-4261){\circle{150}}
\put(6302,-4261){\circle{150}}
\put(6002,-4261){\circle{150}}
\put(5703,-1861){\framebox(2100,2100){}}
\put(5703,-4861){\framebox(2100,2100){}}
\put(5853,314){\makebox(0,0)[lb]{\smash{$S_1$}}}
\put(5853,-2686){\makebox(0,0)[lb]{\smash{$S_2$}}}
\put(6678,-2386){\makebox(0,0)[lb]{\smash{$\times$}}}
\put(11701,-1111){\line( 1, 0){975}}
\put(11701,-1186){\line( 1, 0){975}}
\put(11701,-1261){\line( 1, 0){975}}
\put(11701,-1336){\line( 1, 0){975}}
\put(11701,-1411){\line( 1, 0){975}}
\put(11701,-1486){\line( 1, 0){975}}
\put(11701,-1561){\line( 1, 0){975}}
\put(11701,-1636){\line( 1, 0){975}}
\put(11701,-1711){\line( 1, 0){975}}
\put(11701,-3962){\line( 1, 0){975}}
\put(11701,-4037){\line( 1, 0){975}}
\put(11701,-4112){\line( 1, 0){975}}
\put(11701,-4187){\line( 1, 0){975}}
\put(11701,-4262){\line( 1, 0){975}}
\put(11701,-4337){\line( 1, 0){975}}
\put(11701,-4412){\line( 1, 0){975}}
\put(11701,-4487){\line( 1, 0){975}}
\put(11701,-4562){\line( 1, 0){975}}
\put(11152,-3192){\circle*{50}}
\put(11287,-3312){\circle*{50}}
\put(11152,-3597){\circle*{50}}
\put(11617,-3372){\circle*{50}}
\put(11812,-3237){\circle*{50}}
\put(11917,-3477){\circle*{50}}
\put(12187,-3267){\circle*{50}}
\put(12097,-3057){\circle*{50}}
\put(12592,-3387){\circle*{50}}
\put(12397,-3147){\circle*{50}}
\put(11268, 14){\circle*{50}}
\put(11133,-121){\circle*{50}}
\put(11268,-241){\circle*{50}}
\put(11133,-526){\circle*{50}}
\put(11460,-504){\circle*{50}}
\put(11598,-301){\circle*{50}}
\put(11793,-166){\circle*{50}}
\put(11898,-406){\circle*{50}}
\put(12168,-196){\circle*{50}}
\put(12078, 14){\circle*{50}}
\put(11898,-706){\circle*{50}}
\put(11538,-781){\circle*{50}}
\put(11778,-946){\circle*{50}}
\put(11433,-1246){\circle*{50}}
\put(12378,-76){\circle*{50}}
\put(11073, 74){\circle*{50}}
\put(11523,-3646){\circle*{50}}
\put(11478,-3061){\circle*{50}}
\put(12303,-3511){\circle*{50}}
\put(11402,-3511){\circle*{50}}
\put(11852,-3061){\circle*{50}}
\put(12077,-3661){\circle*{50}}
\put(3076,-2311){\vector( 1, 0){2250}}
\put(8251,-2311){\vector( 1, 0){2250}}
\put(10803,-1861){\framebox(2100,2100){}}
\put(10803,-4861){\framebox(2100,2100){}}
\put(10953,314){\makebox(0,0)[lb]{\smash{$S_1$}}}
\put(10953,-2686){\makebox(0,0)[lb]{\smash{$S_2$}}}
\put(11778,-2386){\makebox(0,0)[lb]{\smash{$\times$}}}
\put(11178,-5300){\makebox(0,0)[lb]{\smash{smooth}}}
\put(11178,-5800){\makebox(0,0)[lb]{\smash{dual to $X_{U(1)}$}}}
\put(5800,-5300){\makebox(0,0)[lb]{\smash{singular}}}
\put(5800,-5800){\makebox(0,0)[lb]{\smash{dual to $X^\sharp$}}}
\put(700,-5300){\makebox(0,0)[lb]{\smash{smooth}}}
\put(700,-5800){\makebox(0,0)[lb]{\smash{dual to $X_{G=0}$}}}
\put(13051,-1636){\makebox(0,0)[lb]{\smash{$G$-flux}}}
\put(13051,-4336){\makebox(0,0)[lb]{\smash{$G$-flux}}}
\end{picture}
$$
\caption{A Brane to Flux Transition} \label{fig:btof}
\end{figure}
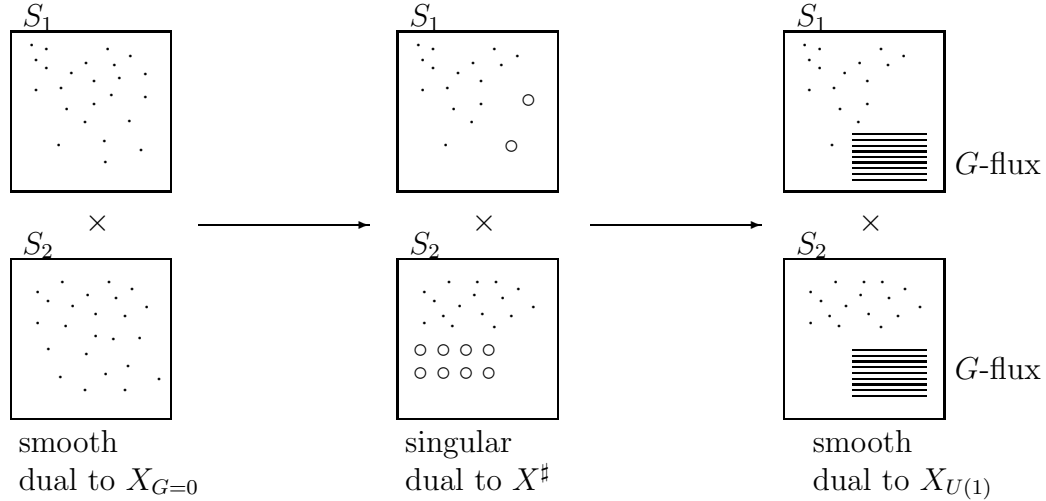

This completes the proof that we have obtained the required extremal
transition. One begins with the smooth $S_1\times S_2$ with 24
M2-branes dual to $X_{G=0}$. One then lets $S_1$ acquire two
$\C^2/\Z_2$ singularities and we push 4 M2-branes into each of these
singularities. This produces degrees of freedom over $S_2$
corresponding to an $(\SU(2)\times\SU(2))\rtimes\Z_2$-bundle. By
making $S_2$ an orbifold $\KIII/\Z_2$ we make the monodromy of this
bundle exchange the $\SU(2)$ factors. This produces a model dual to
$X_1$ before it is resolved. Now we may allow all the singularities to
be resolved resulting in a smooth $S_1\times S_2$ with a $G$-flux and
only 16 M2-branes. We depict this transition in figure \ref{fig:btof},
where dots represent ``free'' point-like instantons, and circles
represent $\C^2/\Z_2$ singularities (containing point-like instantons).

Some subtleties of the geometry of this kind of transition were
discussed in \cite{KMP:enhg}. The curve of type $\mathrm{III}$ fibres
in $X^\sharp$ is a curve of genus 7. This means that F-theory on
$X^\sharp$ will have 7 hypermultiplets in the adjoint representation
of the corresponding enhanced $\SU(2)$ gauge symmetry. Part of the
resolution to $X_{\GU(1)}$ corresponds to giving vacuum expectation
values to these hypermultiplets which breaks the $\SU(2)$ to the
observed $\GU(1)$ gauge group. 

These 7 hypermultiplets also account for the following. The precise
form of $X_{\GU(1)}$ given by (\ref{eq:X1E7}) has a curve of
$\mathrm{III}$ fibres and therefore might be associated with an
$\SU(2)$ gauge group. Moving to a generic complex structure this curve
of fibres is replaced by more generic $\mathrm{I}_1$'s. Thus, when the
complex structure takes the form (\ref{eq:X1E7}) there is a divisor
of the form $\P^1\times C$ for a curve $C$ of genus 7. A generic
deformation turns this divisor into a collection of 12 disjoint
rational curves as described in \cite{KMP:enhg}. This is the geometric
picture of the $\SU(2)$ gauge group being Higgs'ed to a $\GU(1)$
subgroup. 

%%%%%%%%%%%%%

\section{Nonperturbative Effects} \label{s:nonp}

So far we have skated around the boundary of the moduli space that is
accessible from F-theory techniques. Now we will try to make a couple
of general statements about our compactifications regarding the deep
interior of the moduli space.

\subsection{Volume-obstructing $G$-fluxes} \label{ss:vol}

All of the $G$-fluxes discussed so far do not obstruct an overall
metric rescaling of either K3 surface $S_1$ or $S_2$. The supergravity
analysis of $G$-fluxes \cite{BB:8} necessarily happens at the large
volume limit and thus is unlikely to observe any other kind of
obstruction. In this section we will look for $G$-fluxes which do
obstruct such dilatations.

In the previous section we restricted attention to \CY\ threefolds
which are elliptic and K3 fibrations. This is an artificial
restriction and there most certainly exist extremal transitions in and
out of this class.

A simple example can be borrowed from \cite{CGH:con}. Consider the
intersection $X_1$ of two hypersurfaces
\begin{equation}
\begin{split}
  (x_1^4 + x_3^4 - x_4^4)y_0 + (x_0^4+x_2^4+x_4^4)y_1&=0\\
  x_1y_1 + x_2y_2 &=0,
\end{split}
\end{equation}
in $\P^4\times\P^1$. A projection onto the $\P^1$ factor manifestly
gives $X_1$ the structure of a K3 fibration. This fibration has 16
$\P^1$'s corresponding to sections. We may contract these $\P^1$'s to
produce 16 nodes in $X_1$ which may then be deformed to produce the
quintic threefold \cite{CGH:con}.

The type IIA string compactified on $X_1$ is dual to some $E_8\times
E_8$ heterotic string compactified on $\KIII\times T^2$. It is true
that, since $X_1$ is not an elliptic fibration with a section, one is
not free to vary the size of the heterotic string's $T^2$. It will
be the kind of heterotic string compactification considered in section
4.2 of \cite{KV:N=2}. To find precisely the heterotic dual one would
find an extremal transition to an elliptic fibration with a section,
use the F-theory rules we used above, and then follow the transition
back. As the details will not concern us, we choose not to perform
this exercise and just note that a solution exists.

The vector multiplet moduli space, $\cM_V$, of the type IIA string
compactified on $X_1$ is determined in the usual way from the
prepotential $\cF$. To leading order in $\alpha'$, this prepotential is
cubic. If this were the case exactly, the moduli space would be
locally of the
form
\begin{equation}
\frac{\GO(2,n-1)}{\GO(2)\times\GO(n-1)} \times \frac{\Sl(2,\R)}{\GU(1)},
    \label{eq:sK}
\end{equation}
where $n=h^{1,1}(X_1)$. We refer to \cite{me:tasi99} for more
details. The second factor of (\ref{eq:sK}) corresponds to $B+iJ$
integrated over a section of the K3 fibration. That is, it controls
the size of the section. The first factor corresponds to $B+iJ$
integrated over the monodromy-invariant part of $H_2$ of the K3
fibres.

Upon compactification on a circle to three dimensions, $\cM_V$ is
elevated to the quaternionic K\"ahler manifold $\cM_1$. This proceeds
via the so-called ``$c$-map'' of \cite{CFG:II}. Ignoring all
nonperturbative effects, we may model this via a parabolic subgroup
decomposition 
\begin{equation}
\begin{split}
\cM_1 &=
\frac{\GO(4,n+1)}{\GO(4)\times\GO(n+1)}\\
&=
\frac{\GO(2,n-1)}{\GO(2)\times\GO(n-1)}\times
\frac{\GO(2,2)}{\GO(2)\times\GO(2)}\times\R^{2n+2}\\
&=
\frac{\GO(2,n-1)}{\GO(2)\times\GO(n-1)}\times
\frac{\Sl(2,\R)}{\GU(1)}\times\frac{\Sl(2,\R)}{\GU(1)}\times\R^{2n+2}.
\end{split}  \label{eq:qK}
\end{equation}
The last factor corresponds to the Wilson lines (with the vector
dualized to a scalar) of the $n+1$ $\GU(1)$'s we had in four
dimensions. The latter $\Sl(2,\R)/\GU(1)$ contains degrees of freedom
associated to the radius of the circle and the $B$-field.

Now, the important thing to notice is that there are world-sheet
instanton corrections to the prepotential as seen, for example, in
\cite{KV:N=2}. The instantons generically arise from instantons wrapped
around the sections of the K3-fibration. This is one way of seeing
that the size of the section is dual to the coupling of the
dual heterotic string.

Thus, if we follow $X_1$ through the conifold transition described
above, we necessarily enter into the deep ``strongly-coupled'' part of
the moduli space away from the limit where it looks like the symmetric
space (\ref{eq:qK}). The compactification on the quintic threefold is
therefore ``stuck'' in this region as the degree of freedom
corresponding to the size of the section is lost. 

What is the ``strongly-coupled'' region of the moduli space in the
language of M-theory on $S_1\times S_2$? The nonperturbative effects
contributing to the moduli space metric can only arise as divisors
with holomorphic Euler characteristic 2 on $S_1\times S_2$ following
the arguments of \cite{Wit:super}. Such divisors correspond to
$\KIII\times\P^1$ (or, obviously, $\P^1\times\KIII$). The ``coupling''
associated to the moduli space $\cM_1$ is therefore determined by
$\Vol(S_1)\times \Vol(S_2)^2$.

We may derive the (mirror of the) latter result in another
way. M-theory on $S_1\times S_2$ is dual to the heterotic string on
$T^3\times S_2$. The moduli space of the heterotic string on $S_2$
will contain a factor which looks something like $O(4,20)/(O(4)\times
O(20))$ when the K3 surface $S_2$ is large. Worldsheet instanton
corrections wrapping $\P^1$'s inside $S_2$ will warp this moduli space
along the lines of \cite{W:K3inst}. The coupling is therefore measured
by the volume of $S_2$, {\em as measured by the heterotic string}. The
relationship between the metric of the heterotic string and the metric
on M-theory was determined in \cite{W:dyn}. The result is that the
volume of $S_2$ as measured by the heterotic string is replaced by
$\Vol(S_1)^2\Vol(S_2)$.

The natural claim, therefore, is that when we pass through a conifold
transition, as above, that kills the K3 fibration structure, we force
$\Vol(S_1)\Vol(S_2)^2$ to be fixed. That is, we lose the moduli which
allow us to make $S_1$ and $S_2$ simultaneously large. This
necessarily takes us away from the supergravity analysis of $G$-flux. 

Given the $\GO(\Gamma_{4,20})$ T-duality of K3 surfaces (at least for
the spin connection embedded in the gauge group) the obvious thing to
conjecture is that we have a $G$-flux similar to that considered in
section \ref{ss:K3sing}, where $\omega_1$ has picked up some 0-form or
4-form component. Forcing the 4-plane $\Pi$ to be perpendicular to
$\omega_1$ would now force the volume of $S_1$ (corrected to
$\Vol(S_1)\Vol(S_2)^2$ as above) to be fixed. This is rather like the
way that type IIA strings compactified on a K3 surface may pick up
nonabelian gauge groups associated with the finite size of the K3
surface, rather than singularities in the K3 surface \cite{me:lK3}.

Adding a 0-form or a 4-form to $\omega_1$ clearly violates the
primitivity condition on $G$. This should not concern us greatly
however since we expect the supergravity analysis to fail.

It is difficult to be more quantitative at this stage because we are
necessarily dealing with regions of the quaternionic K\"ahler moduli
spaces where the nonperturbative effects are strong.

Finally, suppose we can find a \CY\ $Z$ in the web of possibilities
such that neither $Z$ nor the mirror of $Z$ is a K3 fibration. This
would mean that both $\Vol(S_1)\Vol(S_2)^2$ and $\Vol(S_1)^2\Vol(S_2)$
are fixed. That is, each K3 surface has fixed size. 

Given the statistics of K3 fibrations \cite{AKMS:k3} it is surely
likely that such a \CY\ threefold $Z$ exists. A proof along the lines of
\cite{AKMS:k3} might be a little difficult however. Just because a
\CY\ threefold contains a K3 surface does not mean that this K3
surface need be compatible with the torus actions considered in
\cite{AKMS:k3}. Thus, one may have to go beyond toric methods to prove
the non-existence of the K3 fibration.

\subsection{Even more transitions} \label{ss:trans}

So far we have just looked at the geometry of the \CY\ threefold $X$
for M-theory compactified on $X\times T^2$. Might it be that
the 2-torus can also play a r\^ole to yield more possibilities?

Consider first the five-dimensional theory obtained from M-theory on a
\CY\ threefold $X$. Let $X$ be a K3 fibration without reducible
fibres. The moduli space of vector multiplets will generically be of
some form
\begin{equation}
  \cM_{5,V}=\frac{\GO(1,n-2)}{\GO(n-2)} \times \R_+,
\end{equation}
where $n=h^{1,1}(X)$. This moduli space may be viewed as the classical
K\"ahler cone of the K3 fibre (with normalized volume) times a factor
of $\R_+$ for the size of the base. Generically the Picard number of
the K3 fibre will be given by the dimension of the monodromy-invariant
part of $H^2$ of the fibres. Note that the real dimension of
$\cM_{5,V}$ is $h^{1,1}(X)-1$ since one of the deformations of
K\"ahler form (the overall volume of $X$) defects to the
hypermultiplet moduli space.

M-theory on $X$ may acquire an enhanced nonabelian gauge symmetry in
the usual way if the K3 fibres acquire an ADE singularity.  Let us
denote by $\Gamma_{1,n-2}$ the Picard lattice of the K3 fibre. It should
then be clear that enhanced gauge symmetry corresponds to viewing
$\cM_{5,V}$ as $\Gr(\Gamma_{1,n-2})$ and letting the space-like
1-plane be perpendicular to at least one vector of length squared $-2$
in $\Gamma_{1,n-2}$. Many extremal transitions of $X$ proceed through
such enhanced gauge symmetry points in the moduli space.

Now consider M-theory on $X\times S^1$ or, in other words, the type IIA string
on $X$. Now the vector multiplet moduli space becomes (ignoring
instanton effects)
\begin{equation}
  \cM_{6,V}=\frac{\GO(2,n-1)}{\GO(2)\times\GO(n-1)} \times
  \frac{\Sl(2,\R)}{\GU(1)}.
\end{equation}
The first factor can be viewed as $\Gr(\Gamma_{2,n-1})$ where
$\Gamma_{2,n-1}$ is the Picard lattice of the K3 fibre plus $U$, where
$U$ is generated by $H^0$ and $H^4$ of the fibre. Now we may have
enhanced gauge symmetries whenever the corresponding space-like
2-plane is perpendicular to a vector in $\Gamma_{2,n-1}$ of length
squared $-2$. This means, that in addition to enhanced gauge groups
occurring whenever the K3 fibre acquires a singularity, one may also
get an enhancement if the K3 surface acquires a special overall
volume. This is exactly how the $\SU(2)$ gauge group arises in the
example of section 4.2 of \cite{KV:N=2}.

Now go to M-theory on $X\times T^2$. Now the vector multiplet moduli
space becomes (ignoring instanton effects)
\begin{equation}
  \cM_{6,V}=\frac{\GO(4,n+1)}{\GO(4)\times\GO(n+1)}.
\end{equation}
We will identify this as $\Gr(\Gamma_{4,n+1})$. Again we expect to see
an enhanced gauge symmetry if the space-like 4-plane is perpendicular
to a vector of length squared $-2$ in $\Gamma_{4,n+1}$. The lattice
$\Gr(\Gamma_{4,n+1})$ should be viewed as one $U$-duality extended
version of $H^{\textrm(even)}(\KIII,\Z)$ as in
\cite{HT:unity,AM:Ud}. These new gauge symmetries, not visible in four
or five dimensional compactifications, cannot be associated with the
\CY\ threefold $X$ alone.

For M-theory on $\KIII\times\KIII$, the identification of
$\Gr(\Gamma_{4,n+1})$ is manifest --- we associate it with one of the K3
surfaces. Even though M-theory has no $B$-field, we always get a
factor looking something like this associated to each K3 surface as we
saw earlier in section \ref{s:CYK}.

This is all evidence for the following claim. If the K3 surface $S_1$
is such that the type IIA string compactified on $S_1$ yields an
enhance gauge symmetry group, then an M-theory compactification on
$S_1\times S_2$ will see the same effect. That is, we obtain a gauge
symmetry from the $S_1$ part but this may be broken by some bundle
over $S_2$.

One may also argue for this proposal as follows. It was shown in
\cite{AM:po} that the $E_8\times E_8$ heterotic string compactified on
a K3 surface with singularity $\C^2/H$ will always produce a
nonperturbative gauge group enhancement by a group containing the
associated $\cH$ so long as enough point-like instantons are embedded
in the singularity.\footnote{From our discussion in section
\ref{ss:K3sing} and the mirror symmetry analysis of \cite{PR:hetmir}
this number of point-like instantons is equal to $c_2$ for a principle
$\cH$-bundle. Thus puts a weak bound on this number of at least
$\dim(\cH)/h^\vee$ where $h^\vee$ is the dual Coxeter number of
$\cH$.}  We simply want to extend this naturally into the moduli space
so that if a K3 surface is of a special volume that leads to enhanced
gauge symmetries, then the heterotic string can also produce the
enhanced symmetries so long as we deal with point-like instantons
accordingly.

There is only one significant objection that can be made to the
structure conjectured above. That is, we have not taken any instanton
effects into account and the moduli spaces will not be of the
Grassmannian form we used. In other words, the T-duality group of the
K3 surface will be broken along the lines of \cite{AP:T}. Therefore it
might be most unfair to treat the $H^0$ and $H^4$ directions in the
cohomology the same as the $H^2$ directions.

We wish only to assert that there will be extremal transitions
associated to these new enhanced gauge symmetries. While it is true
that nonperturbative effects can break gauge symmetries as in
\cite{SW:I}, the extremal transitions (i.e., Higgs--Coulomb
transitions) are not removed by these corrections.

We conjecture, therefore, that there are extremal transitions
associated to M-theory on $X\times T^2$, that are not associated to
extremal transitions on the \CY\ threefold $X$ itself. The extra
structure arising from the $T^2$ part of the compactification allow
for these new transitions.

This means that there are even more possibilities for choices of flux
and M2-branes for M-theory on $\KIII\times\KIII$ than there are \CY\
threefolds!

%%%%%%%%%%%%%%%%%%%%%%%%%%%%%%%%%%%%%%%%%%%%%%%%%%%%%%%%%%%%%%%%%%%

\section*{Acknowledgments}

I wish to thank M.~Douglas, S.~Kachru, R.~Kallosh and D.~Morrison for
useful conversations. This work is supported in part by NSF grant
DMS-0301476, Stanford University, SLAC and the Packard Foundation.

%\bibliographystyle{my-phys}
%\bibliography{string}

\end{document}